\documentclass[conference]{IEEEtran}
\IEEEoverridecommandlockouts
\usepackage{cite}
\usepackage{amsmath,amssymb,amsfonts}
\usepackage{algorithmic}
\usepackage{graphicx}
\usepackage{textcomp}
\usepackage{xcolor}
\usepackage{float}
\usepackage[ruled]{algorithm2e}

\def\BibTeX{{\rm B\kern-.05em{\sc i\kern-.025em b}\kern-.08em T\kern-.1667em\lower.7ex\hbox{E}\kern-.125emX}}

\def\BibTeX{{\rm B\kern-.05em{\sc i\kern-.025em b}\kern-.08em T\kern-.1667em\lower.7ex\hbox{E}\kern-.125emX}}

\newcommand{\mymarginpar}[1]{\marginpar{#1}}
\renewcommand{\marginpar}[1]{}

\newcommand{\ls}[1]{\dimen 0= \fontdimen 6 \the \font
	\lineskip =#1 \dimen0
	\advance \lineskip .5 \fontdimen 5 \the \font
	\advance \lineskip - \dimen 0
	\lineskiplimit =.9 \lineskip
	\baselineskip = \lineskip
	\advance \baselineskip \dimen 0
	\normallineskip \lineskip
	\normallineskiplimit \lineskiplimit
	\normalbaselineskip \baselineskip
	\ignorespaces
}

\newcommand{\bearn}{\begin{eqnarray*}}
	\newcommand{\eearn}{\end{eqnarray*}}
\newcommand{\barr}{\begin{array}}
	\newcommand{\earr}{\end{array}}

\newcommand{\N}{{\cal N}}

\newtheorem{definition}{Definition}
\newtheorem{assumption}[definition]{Assumption}
\newtheorem{property}[definition]{Property}
\newtheorem{proposition}[definition]{Proposition}
\newtheorem{lemma}[definition]{Lemma}
\newtheorem{theorem}[definition]{Theorem}
\newtheorem{corollary}[definition]{Corollary}
\newtheorem{example}[definition]{Example}
\newtheorem{remark}[definition]{Remark}

\newcommand{\benum}{\begin{enumerate}}
	\newcommand{\eenum}{\end{enumerate}}

\newcommand{\bdesc}{\begin{description}}
	\newcommand{\edesc}{\end{description}}

\newcommand{\bfig}[2]{
	\begin{figure}\centering\includegraphics[width=#2]{#1}
	}
	\newcommand{\brotatefig}[2]{
		\begin{figure}[htbp]\centerline{
				\epsfig{figure={#1},clip=,angle=-90,width={#2}}
			}
		}
		
		\newcommand{\bfigfirst}[2]{\begin{figure}[h]\centerline{\setlength{\epsfxsize}{#2}\epsffile{#1}}}
			\newcommand{\efig}[2]{\caption{#2}\label{fig:#1}\end{figure}\mymarginpar{fig:#1}}
		\newcommand{\erotatefig}[2]{\caption{#2}\label{fig:#1}\end{figure}\mymarginpar{fig:#1}}
	\newcommand{\rfig}[1]{Figure \ref{fig:#1}}
	\newcommand{\btab}[1]{\begin{table}\centering\begin{tabular}{#1}}
			\newcommand{\etab}[3]{\end{tabular}\caption[#3]{#2}\label{tab:#1}\end{table}\mymarginpar{tab:#1}\vspace{.1in}}

	\newcommand{\btabular}[1]{\begin{center}\begin{tabular}{#1}}
			\newcommand{\etabular}{\end{tabular}\end{center}}
	\newcommand{\bdefin}[1]{\begin{definition}\mymarginpar{def:#1}\label{def:#1}}
		\newcommand{\edefin}{\end{definition}}
	
	\newcommand{\bpro}[1]{\begin{property}\mymarginpar{pro:#1}\label{pro:#1}}
		\newcommand{\epro}{\end{property}}
	
	\newcommand{\bprop}[1]{\begin{proposition}\mymarginpar{prop:#1}\label{prop:#1}}
		\newcommand{\eprop}{\end{proposition}}
	
	\newcommand{\blem}[1]{\begin{lemma}\mymarginpar{lem:#1}\label{lem:#1}}
		\newcommand{\elem}{\end{lemma}}
	\newcommand{\rlem}[1]{Lemma \ref{lem:#1}}
	\newcommand{\bass}[1]{\begin{assumption}\mymarginpar{the:#1}\label{ass:#1}}
		\newcommand{\eass}{\end{assumption}}
	
	\newcommand{\bthe}[1]{\begin{theorem}\mymarginpar{the:#1}\label{the:#1}}
		\newcommand{\ethe}{\end{theorem}}
	\newcommand{\rthe}[1]{Theorem \ref{the:#1}}

	\newcommand{\bcor}[1]{\begin{corollary}\mymarginpar{cor:#1}\label{cor:#1}}
		\newcommand{\ecor}{\end{corollary}}
	
	\newcommand{\bax}[1]{\begin{axiom}\mymarginpar{ax:#1}\label{ax:#1}}
		\newcommand{\eax}{\vspace{-.1in} \end{axiom}}
	
	\newcommand{\bex}[2]{\vspace{.1in}\begin{example}\mymarginpar{ex:#1}{\bf #2}\label{ex:#1} }
		\newcommand{\eex}{\end{example}\vspace{.3cm}}
	
	\newcommand{\brem}[1]{\begin{remark}\mymarginpar{rem:#1}\label{rem:#1}\em}
		\newcommand{\erem}{\end{remark}}
	
	\newcommand{\beq}[1]{\mymarginpar{eq:#1}\begin{equation}\label{eq:#1}}
	\newcommand{\beqno}[1]{\mymarginpar{eq:#1}\begin{eqnarray}\nonumber}
	\newcommand{\eeq}{\end{equation}}
	\newcommand{\eeqno}{&&\end{eqnarray}}
\newcommand{\req}[1]{(\ref{eq:#1})}

\newcommand{\bear}[1]{\mymarginpar{eq:#1}\begin{eqnarray}\label{eq:#1}}
\newcommand{\bearno}[1]{\mymarginpar{eq:#1}\begin{eqnarray}\nonumber}
\newcommand{\eear}{\end{eqnarray}}
\newcommand{\eearno}{\end{eqnarray}}
\usepackage{environ}
\NewEnviron{es}{%
\begin{equation}\begin{split}
\BODY
\end{split}\end{equation}
}
\newcommand{\bsel}{\left\{\begin{array}{cl}}
\newcommand{\esel}{\end{array}\right.}
\newcommand{\bmat}[1]{\left[\begin{array}{#1}}
\newcommand{\emat}{\end{array}\right]}
\newcommand{\bsec}[2]{\mymarginpar{sec:#2}\section{#1}\label{sec:#2}}
\newcommand{\rsec}[1]{Section \ref{sec:#1}}

\newcommand{\bsubsec}[2]{\mymarginpar{subsec:#2}\subsection{#1}\label{subsec:#2}}

\newcommand{\bapp}{\begin{appendices}}
\newcommand{\eapp}{\end{appendices}}
\def\R{I\kern-0.30em R}
\def\N{I\kern-0.30em N}
\def\P{I\kern-0.30em P}

\begin{document}

\title{Percolation Threshold for Competitive Influence in Random Networks}

\author{Yu-Hsien Peng, Ping-En Lu, Cheng-Shang Chang and Duan-Shin Lee\\
Institute of Communications Engineering, National Tsing Hua University\\
Hsinchu 30013, Taiwan, R.O.C.\\
Email: znb907512520@gmail.com; j94223@gmail.com; cschang@ee.nthu.edu.tw; lds@cs.nthu.edu.tw\\
}

\maketitle

\begin{abstract}
In this paper, we propose a new averaging model for modeling the competitive influence of $K$ candidates among $n$ voters in an election process. For such an influence propagation model, we address the question of how many seeded voters a candidate needs to place among undecided voters in order to win an election. We show that for a random network generated from the stochastic block model, there exists a percolation threshold for a candidate to win the election if the number of seeded voters placed by the candidate exceeds the threshold. By conducting extensive experiments, we show that our theoretical percolation thresholds are very close to those obtained from simulations for random networks and the errors are within $10\%$ for a real-world network.
\end{abstract}


\bsec{Introduction}{introduction}

Due to the advent of online social networks, such as Twitter, Facebook, it becomes possible to spread the information/misinformation to influence people in a short period of time. Studying {\em opinion dynamics} to understand how opinions are propagated through social networks is of importance in social network analysis. In particular, Kempe, Kleinberg, and Tardos \cite{kempe2003maximizing} proposed two basic models for influence propagation of a single idea (source, product) in a social network: the Independent Cascade (IC) model and the Linear Threshold (LT) model. In the IC model, a susceptible node is activated through one of its neighboring node with a certain influence probability. On the other hand, a susceptible node in the LT model is activated if the sum of the influences of its neighbors exceeds a certain threshold. Instead of focusing on the propagation of a single idea in social networks, there are various extensions of the IC model and the LT model for multiple competing ideas (see, e.g., \cite{bharathi2007competitive, carnes2007maximizing,kostka2008word, he2012influence, shirazipourazad2012influence, lin2015learning, lin2015analyzing, li2015getreal, tong2018misinformation, tong2018distributed}).

Both the IC model and the LT model are {\em exclusive} in the sense that an activated node will not change its state once it is activated. Though such exclusive influence propagation models might be suitable for modeling the purchase of a product, it may not be appropriate for modeling an election process, where the opinions of voters might change with respect to time. As discussed in \cite{das2014modeling}, there are several non-exclusive models in the literature that could be used for modeling an election process, including the averaging model (the DeGroot model \cite{degroot1974reaching}), the bounded confidence model (the HK model \cite{hegselmann2002opinion}), and the voter model \cite{clifford1973model, holley1975ergodic}. However, these influence propagation models were originally designed for positive influence (edge weights) only. Recent works in \cite{proskurnikov2016opinion, dhamal2018two} showed that the influence could also be negative. A negative influence (edge weight) between two neighboring voters implies that these two voters might be from two hostile camps and tend to adopt opposite opinions. Negative influence poses a technical challenge for the analysis of opinion dynamics as the opinions of voters might not be bounded and thus need to be renormalized.

To tackle such a problem, in this paper we propose a new averaging model for modeling the competitive influence of $K$ candidates among $n$ voters in an election process. We assume that each voter (in his/her mind) has a $K$-dimensional probability preference vector (PPV) that indicates the preference of a voter on the $K$ candidates. The opinion dynamic of a voter then consists of two steps: (i) the combined influence on a voter is computed by averaging over the weighted PPVs of its neighbors, and (ii) the PPV of a voter is then updated and renormalized by a softmax decision based on the combined influence. For such a model, we pose the question of how many seeded voters (voters who are stubborn and will not change their mind) a candidate needs to place among undecided voters in order to win over the votes from undecided voters. We address such a question by analyzing our opinion dynamic model in a random network generated from the stochastic block model. Inspired by the percolation analysis for the (single-idea) influence maximization problem in \cite{morone2015influence}, we show that (under certain technical conditions) there exists a percolation threshold for a candidate to win the election if the number of seeded voters placed by the candidate exceeds the threshold. To the best of our knowledge, our percolation results seem to be the first one in the competitive influence maximization problem. By conducting extensive simulations, we show that our theoretical percolation thresholds are very close to those obtained from simulations. For the real-world network, {\em Political Blogs} in \cite{adamic2005political}, the errors are found to be less than $10\%$.
Additional experimental results for several real-world networks, including the Youtube social network \cite{yang2015defining} and the email network \cite{yin2017local, leskovec2007graph}, also show the percolation phenomenon.

The rest of the paper is organized as follows. In \rsec{system}, we introduce the system model, including the model for competitive influence propagation model and the model for the influence in a network. We then analyze our competitive influence propagation model for random networks generated by the stochastic block models in \rsec{percolation}. Various experiments are conducted in \rsec{exp} to verify the percolation phenomenon in both random networks and several real-world networks. The paper is then concluded in \rsec{con}, where we discuss possible extensions of our work.

In Table \ref{table1}, we provide a list of notations that are used in this paper.
\begin{table}[tb]
 	\caption{List of notations}
 	\begin{center}
 		\begin{tabular}{|l|l|}
 			\hline
 			\textbf{}&{\textbf{Description}}\\
 			\hline
 			$n$ & The total number of voters (nodes)\\
 			$K$ & The total number of candidates\\
 			$q(u,w)$ & The influence from voter $u$ to voter $w$\\
 			$Q$ & $Q=(q(u,w))$ the influence matrix\\
 			$h_{u,k}(t)$ & The preference probability of voter $u$ for candidate $k$\\
 			& at time $t$\\
 			$h_u(t)$ & The preference probability vector (PPV) of voter $u$ at time $t$\\
 			$h_k$ & The initial preference probability of an undecided voter\\
 			& for candidate $k$\\
 			$z_{u,k}(t)$ & The combined influence on voter $u$ for candidate $k$ at time $t$\\
 			$S_k$ & The set of seeded voters for candidate $k$\\
 			$a_{u,w}$ & The indicator variable for an edge between $u$ and $w$\\
 			$A$ & $A=(a_{u,w})$ the adjacency matrix of a graph\\
 			$m$ & The total number of edges\\
 			$\beta$ & a parameter for modeling generalized modularity\\
 			$b$ & The total number of blocks in a SBM\\
 			$k_v$ & The degree of node $v$\\
 			$\rho_i$ & The proportion of nodes in block $i$\\
 			$p_{in}$ & The intra-block edge probability\\
 			$p_{out}$ & The inter-block edge probability\\
 			$B_k$ & The set of (seeded) voters in block $k$ (for candidate $k$)\\
 			$B_{b,k}$& The set of seeded voters in block $b$ for candidate $k$\\
 			$\rho_{b,k}$ & The fraction of nodes in $B_{b,k}$\\
 			$B^u_b$ & The set of undecided voters\\
			$\rho^u_{b}$ & The fraction of undecided voters\\
			$\lambda_i$ & the (normalized) degree of a node in block $i$ in \req{SBM3355}\\
 			\hline
 		\end{tabular}
 		\label{table1}
 	\end{center}
\end{table}

\bsec{The system model}{system}

\bsubsec{The model for competitive influence propagation}{model}

\begin{figure}[b]
	\centering
	\includegraphics[width=0.45\textwidth]{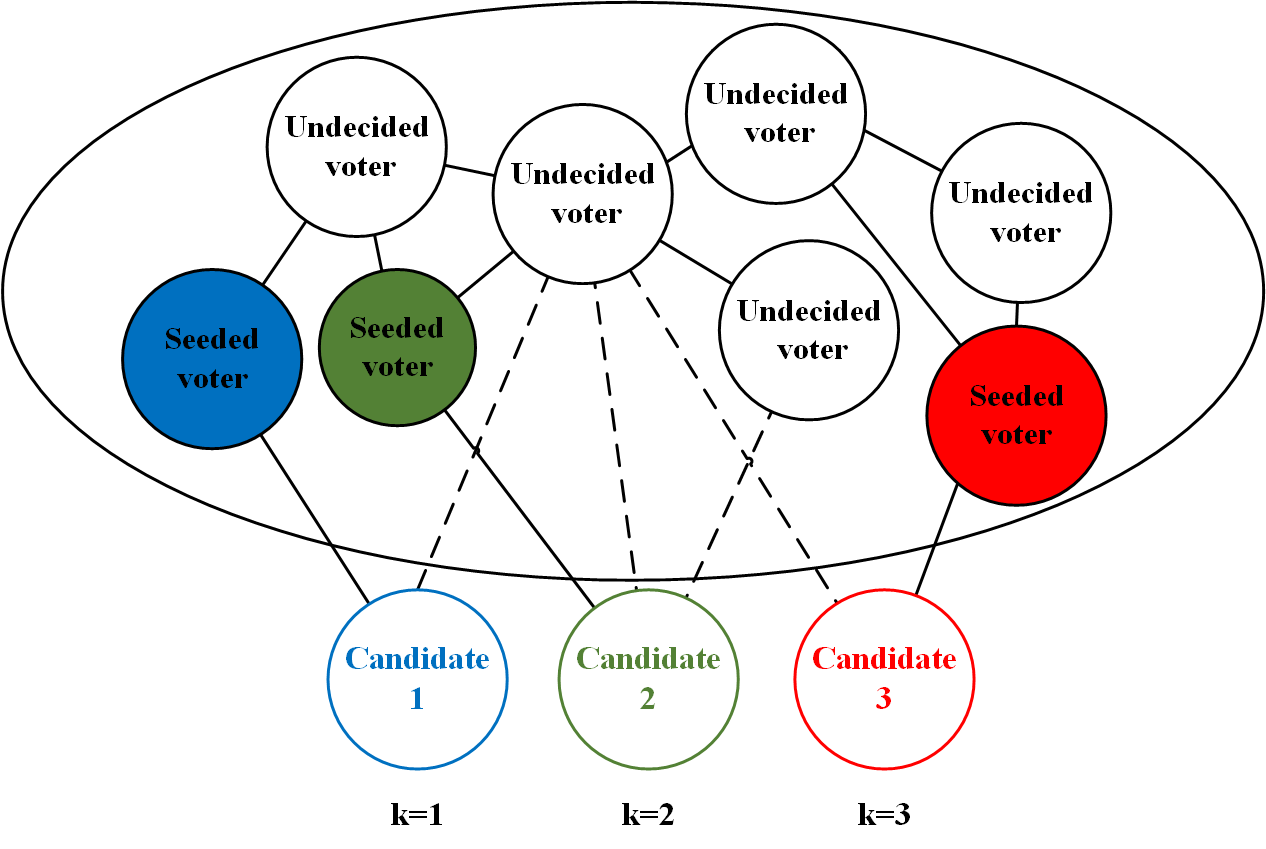}
	\caption{An illustration of our competitive influence model.}
	\label{fig:illustration}
\end{figure}
In this section, we introduce our competitive influence propagation model for modeling an election process during a period of time $T$. In our model, there are $K$ {\em candidates} and $n$ {\em voters} (see \rfig{illustration} for an illustration). At any time $t$, each voter has its own preferences on the $K$ candidates characterized by a {\em probability vector}. Specifically, let $h_{u,k}(t)$ be the preference of voter $u$ for candidate $k$ at time $t$ and $h_u(t)=(h_{u,1}(t), h_{u,2}(t), \ldots, h_{u,K}(t))$ be the preference probability vector (PPV) of voter $u$ at time $t$. In order for $h_u(t)$ to be a probability vector, we need to ensure that $h_{u,k}(t) \ge 0$ for all $k$ and $\sum_{k=1}^K h_{u,k}(t)=1$. In addition to these, we assume that there are two types of voters: {\em seeded voters} and {\em undecided voters}. {\em Seeded voters} are stubborn and their preferences will not be affected by other voters. Thus, their preferences remain the same during the whole election process. As such, the PPV of a seeded voter of candidate $k$ is fixed and set to be $h_{u,k}(t)=1$ and $h_{u,\ell}(t)=0$ for all $\ell \ne k$ and all $0 \le t \le T$. On the other hand, undecided voters' preferences can be influenced by other voters. Moreover, the initial PPVs of undecided voters are set according to a specific probability vector $h=(h_1, h_2, \ldots, h_K)$, i.e., $h_{u,k}(0)=h_k$, $k=1, 2, \ldots, K$, for an undecided voter $u$. The probability vector $h$ could be interpreted as an initial poll for the $K$ candidates at the beginning of the election process.

To model the influence between two voters, we let $q(u,w)$ be the influence from voter $u$ to voter $w$. The $n \times n$ matrix $Q=(q(u,w))$ is called the influence matrix in this paper. The variable $q(u,w)$ is assumed to be real-valued and it is invariant with respect to time. We note that the influence could be asymmetric, i.e., $q(u,w)$ may not be the same as $q(w,u)$. To further model the propagation of the influence, we adopt the widely used influence propagation model in the distributed averaging system in \cite{xiao2007distributed} and the random gossip algorithm in \cite{boyd2006randomized}. Specifically, each undecided voter has a clock which ticks at the times of a Poisson process with rate $1$. Once an undecided voter's clock ticks, it combines the influences from all the other voters. For this, we let $z_{u,k}(t)$ be the combined influence from the other voters on voter $u$ for candidate $k$ at time $t$ and it is computed as follows:
\beq{influence1111}
z_{u,k}(t)=\sum_{w \ne u}q(w,u)h_{w,k}(t).
\eeq
When the clock of voter $u$ ticks at time $t$, voter $u$ first computes the combined influence from the other voters for all the $K$ candidates and then makes a softmax decision \cite{gold1996softmax, bishop2006pattern} to update its preferences on the $K$ candidates at time $t^+$. This is specified by the following update rule:
\beq{influence2222}
h_{u,k}(t^+)=\frac{e^{\theta z_{u,k}(t)} h_{u,k}(t)}{ \sum_{\ell=1}^K e^{\theta z_{u,\ell}(t)} h_{u,\ell}(t) },
\eeq
where $\theta > 0$ is the inverse temperature that characterizes how soft the decision is. Note that if $\theta \to \infty$, then the decision becomes a hard decision, and $h_{u,k}(t^+)=1$ if $k=\mathop{\arg\max}_{1 \le \ell \le K}z_{u,\ell}(t)$ and $0$ otherwise. At the ending time $T$, each voter then votes for the candidate on whom it has the highest preference, i.e., voter $u$ votes for the candidate $k^*$ if $k^*=\mathop{\arg\max}_{1 \le \ell \le K} h_{u,\ell}(T)$ (with ties broken arbitrarily). In our model, one particular view of candidate $K$ is to interpret it as a {\em virtual} candidate, and undecided voters voted for candidate $K$ can be interpreted as undecided voters who do not vote at time $T$. The details of our completive influence propagation model are shown in Algorithm \ref{alg:propagation}, where we denote by $S_k$, $k=1, 2, \ldots, K$, the set of seeded voters selected by the $K$ candidates.

Analogous to the influence maximization in \cite{kempe2003maximizing}, one can also define the competitive influence maximization problem for our model as the problem that asks each candidate to select a set of seeded voters so as to maximize its expected number of votes from undecided voters at the ending time $T$. Such a problem is in general very difficult to solve for a deterministic network. However, as we will show later that there exist very interesting percolation results for {\em random} networks.
\begin{algorithm}[htb]
	\KwIn{The number of voters $n$, the number of candidates $K$, the influence matrix $Q=(q(u, w))$, the inverse temperature $\theta > 0$, the ending time $T$, the $K$ seeded sets, $S_1, S_2, \ldots, S_K$, and the initial PPV at time $0$
		$h = (h_{1}, h_{2}, \ldots, h_{K})$.}
	\KwOut{The PPV of every voter at time $T$
		$h_u(T) = (h_{u, 1}(T), h_{u, 2}(T), \ldots, h_{u,K}(T))$, $u = 1, 2, \ldots, n$. }
	
		\noindent {\bf (1)} Initially, set the voters that are not in any seeded sets as undecided voters and set its PPV by using the initial PPV.
	\\
\noindent {\bf (2)} Generate the clock of each undecided voter by a Poisson process with rate $1$.
\\
	\noindent {\bf (3)} Suppose that the clock of an undecided voter ticks at time $t$.
	\\
	\noindent {\bf (4)} Compute the combined influence on voter $u$ for candidate $k$ at time $t$, i.e., $z_{u, k}(t)$, by using \req{influence1111}.
	\\
	\noindent {\bf (5)} Update the preference of voter $u$ on candidate $k$ at time $t^+$, i.e., $h_{u, k}(t^+)$, by using \req{influence2222}.
	\\
	\noindent {\bf (6)} Repeat from Step 3 until the ending time $T$.
	\\
	\noindent {\bf (7)} For each voter $u$, find $k(u)=\mathop{\arg\max}_{1\leq \ell \leq K} h_{u,\ell}(T)$ and set $h_{u, k(u)}(T)=1$ and $h_{u, \ell}(T)=0$ for $\ell \ne k(u)$.
	\caption{The competitive influence propagation model}
	\label{alg:propagation}
\end{algorithm}

\bsubsec{The model for influence in a network}{influence}

For our competitive influence propagation model, we need a model for the influence matrix $Q=(q(u,w))$. Though such a matrix might be learned from a large dataset of cascades that occurred in a social network (see, e.g., \cite{goyal2010learning, liao2016uncovering}), it is in general very difficult to learn a meaningful influence matrix for a very large network. For our analysis for large networks, we resort to mathematical models. In particular, we choose the generalized modularity of an undirected graph \cite{reichardt2006statistical} as our model for influence. Consider an undirected graph $G=(V,E)$ with the $n \times n$ adjacency matrix $A=(a_{u,w})$, i.e., $a_{u,w}=1$ if there is an edge between node $u$ and node $w$, and $0$ otherwise. Let $m$ be the total number of edges in the undirected graph and $k_u$ be the degree of node $u$ in the graph. Then
\beq{modularity3333}
m=\frac{1}{2}\sum_{u=1}^n\sum_{w=1}^n a_{u,w},
\eeq
and
\beq{modularity4444}
k_u=\sum_{w=1}^n a_{u,w}.
\eeq
The generalized modularity of the graph $G$ is defined as
\beq{modularity5555}
	q(u,w)= \frac{a_{u,w}}{2m}-\beta \frac{k_u}{2m}\frac{k_w}{2m}.
\eeq
If, furthermore, $\beta$ is set to $1$, then it reduced to the original modularity defined in \cite{Newman04}. One intuitive interpretation of the parameter $\beta$ is to view $\beta$ as an index for ``social temperature.'' If $\beta=0$, we note that $q(u,w) > 0$ if $u$ and $w$ are connected by an edge. Thus, there are positive influence between any pairs of two connected nodes. On the other hand, if $\beta=1$, then $q(u,w) < 0$ if $u$ and $w$ are not connected by an edge. Thus, there are negative influence between any pairs of two unconnected nodes. Increasing $\beta$ increases the ``social temperature'' and decreases ``social cohesiveness'' in a network. For the community detection problem, it is well-known (see, e.g., \cite{reichardt2006statistical}) that the parameter $\beta$ can be used a ``resolution'' parameter to detect various time scales of community structure in a network.

One possible generalization of our model for influence is to use the generalized modularity of a {\em sampled graph} \cite{chang2015relative, chang2018probabilistic}. A sampled graph in \cite{chang2015relative, chang2018probabilistic} is obtained by sampling a graph $G=(V,E)$ (with a set of nodes $V$ and a set of edges $E$) according to a specific bivariate distribution $p_{U,W}(u,w)$ that characterizes the probability for the two nodes $u$ and $w$ to appear in the same sample. The marginal distribution $P_{U}(u)= \sum_w p_{U,W}(u,w)$ is the probability that a node $U$ is sampled and it can be used for representing the centrality of a node. The generalized modularity from node $u$ to node $w$ in a sampled graph is defined as
\beq{modularity1111}
	q(u,w)= P_{U,W}(u,w)-\beta P_U(u)P_W(w).
\eeq
There are many known methods to choose the bivariate distribution $p_{U,W}(u,w)$. One commonly used method is the {\em uniform edge sampling}, where $U$ and $W$ are the two ends of a randomly selected edge. In this case, the generalized modularity of a sampled graph in \req{modularity1111} recovers \req{modularity5555} as a special case.

\bsec{Competitive influence propagation in stochastic block models}{percolation}

In this section, we analyze our competitive influence propagation model in random graphs generated by stochastic block models.

\bsubsec{Stochastic block models}{sbm}

We first give a brief introduction of the stochastic block models. The stochastic block model is a generalization of the Erd\"os-R\'enyi random graph \cite{erdos1959random} and it has been widely used for generating random graphs that can be used for benchmarking community detection algorithms (see, e.g., \cite{saade2014spectral, decelle2012mode}). In a stochastic block model with $n$ nodes and $b$ blocks, the $n$ nodes are in general assumed to be evenly distributed to the $b$ blocks. Here we allow the number of nodes in the $b$ blocks to be different. For this, we let $B_i$ be the set of nodes in the $i^{th}$ block and $\rho_i=|B_i|/n$ be the ratio of the number of nodes in the $i^{th}$ block to the total number of nodes. Also, let
\beq{SBM0000}
\rho=(\rho_1, \rho_2, \ldots, \rho_b)
\eeq
be the probability vector that a randomly selected node is in block $i$, $i=1, 2, \ldots, b$. As in the construction of an Erd\"os-R\'enyi random graph, the edges in a random graph from the stochastic block model are generated independently. Specifically, the probability that there is an edge between two nodes within the same block is $p_{in}$ and the probability that there is an edge between two nodes in two different blocks is $p_{out}$. For the ease of our presentation, we denote by $\mbox{SBM}(n,b, p_{in}, p_{out}, \rho)$ a random graph generated from the stochastic block model with $n$ nodes, $b$ blocks, the intra-block edge probability $p_{in}$, the inter-block edge probability $p_{out}$, and $n \rho_i$ nodes in the $i^{th}$ block, $i=1, 2, \ldots, b$.
\begin{figure}[b]
	\centering
	\includegraphics[width=0.45\textwidth]{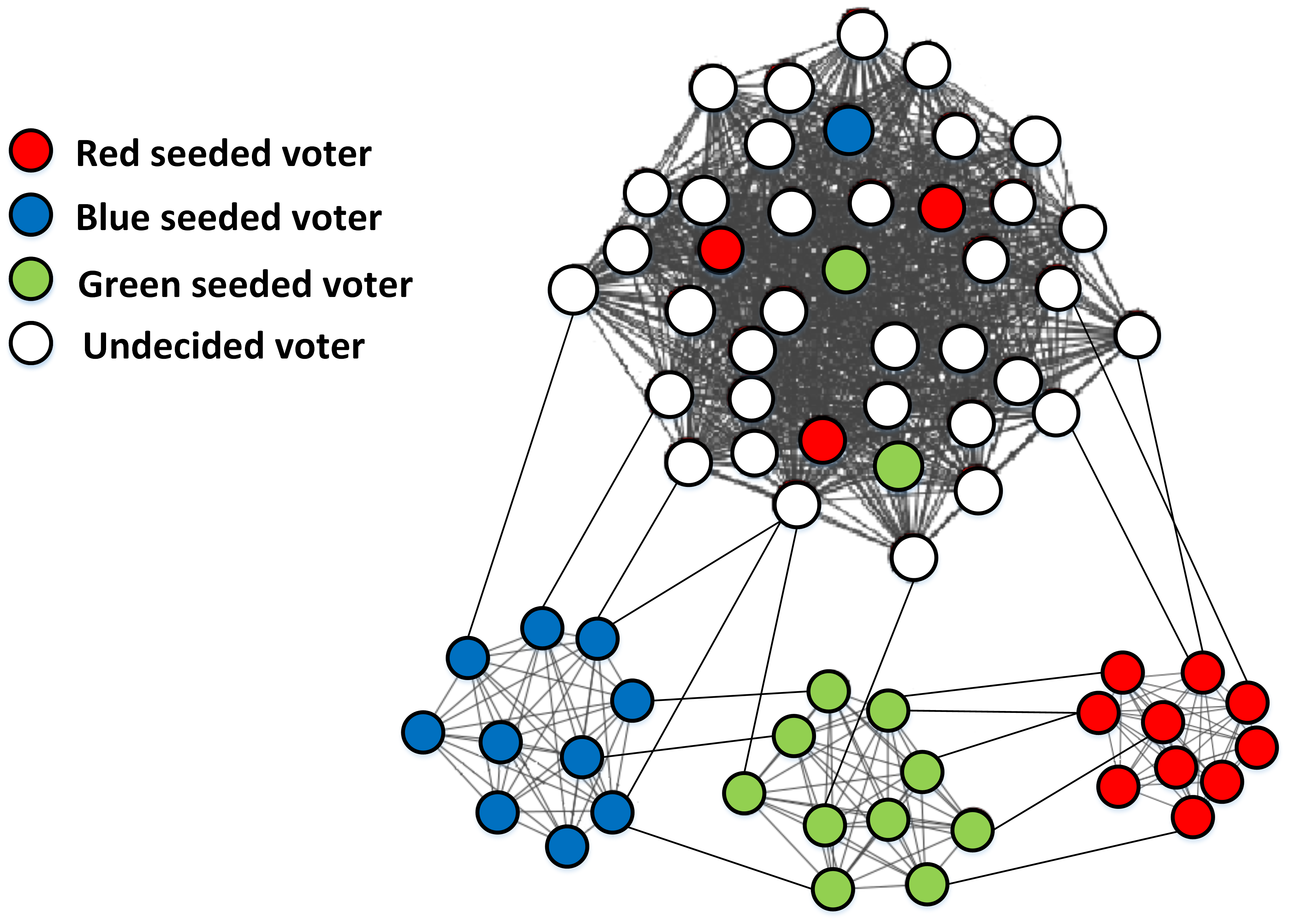}
	\caption{The competitive influence model in the stochastic block model}
	\label{fig:SBM}
\end{figure}

Now we consider the competitive influence propagation model in a random graph generated from the stochastic block model (see \rfig{SBM}). Suppose there are $b-1$ candidates and the nodes in $B_k$ are all seeded voters (basic supporters) for candidate $k$, $k=1, 2, \ldots, b-1$. As such, undecided voters only exist in block $b$. To attract the votes from block $b$, each candidate then (randomly) selects a set of seeded voters in block $b$. Specifically, let $B_{b,k}$ be the set of seeded voters selected by candidate $k$ in block $b$ and $B^u_{b}=B_b \backslash(\cup_{k=1}^{b-1} B_{b,k})$ be the set of undecided voters in block $b$. Let $\rho_{b,k}=|B_{b,k}|/n$ be the ratio of the number of seeded voters in $B_{b,k}$, $k=1, 2, \ldots, b-1$, to the total number of voters, and $\rho^u_{b}={|B^u_b|}/{n}$ be the ratio of the number of undecided voters to the total number of voters. Note that
\beq{SBM7766}
\rho^u_{b}= \rho_b -\sum_{k=1}^{b-1}\rho_{b,k}.
\eeq
For each undecided voter $u$, we assume that its initial PPV is $h$ (as described in Algorithm \ref{alg:propagation}), i.e., $h_{u,k}(0)=h_k$ for $k=1, 2, \ldots, b-1$. On the other hand, for a seeded voter $u$ of candidate $i$, we set $h_{u,k}(0)=\delta_{i,k}$, $k=1, 2, \ldots, b-1$, where $\delta_{i,k}$ is the delta function that has value $1$ if $i=k$ and $0$ otherwise. As mentioned before, seeded voters are stubborn and will not change their PPVs with respect to time. Only undecided voters can be influenced by the other voters.

\bsubsec{Percolation threshold}{percolation}

In this section, we analyze the competitive influence propagation model in a random graph generated from the stochastic block model. We will show that if the number of seeded voters placed by a candidate in the set of undecided voters exceeds a certain threshold, then the candidate is going to win over (almost) all the undecided votes.

In the following lemma, we first derive a mean field approximation for the combined influence on an undecided voter $u$ at time $t$. The mathematical theory behind this mean field approximation is the strong law of large numbers. Due to space limitation, the proof of \rlem{SBM0} is given in in Appendix A.
\blem{SBM0}
For $k=1, 2, \ldots, b-1$, let
\beq{SBM9955}
h_k(t)=\frac{1}{n \rho^u_{b}}\sum_{w \in B^u_b}h_{w,k}(t)
\eeq
be the average preference of undecided voters for candidate $k$. The combined influence on voter $u$ at time $t$ has the following approximation:
\bear{SBM9999}
&&n z_{u,k}(t)\approx \frac{1}{\sum_{\ell=1}^b \rho_\ell \lambda_\ell}\Big (\rho_k p_{out}+\rho_{b,k}p_{in} +{\rho^u_{b} p_{in}}{h_k(t)} \nonumber \\
&&\quad\quad-
\beta \frac{\rho_k \lambda_k \lambda_b+(\rho_{b,k}+{\rho^u_{b}}{h_k(t)}) (\lambda_b)^2}{\sum_{\ell=1}^b \rho_\ell \lambda_\ell} \Big),
\eear
where
\beq{SBM3355}
\lambda_k=\rho_k p_{in}+(1-\rho_k)p_{out}, \;k=1, 2, \ldots, b.
\eeq
\elem

It is interesting to see that the mean field approximation in \req{SBM9999} for the combined influence is independent of $u$ and thus it is the same for all undecided voters. In some sense, undecided voters are well ``mixed'' in random networks as they are subject to the same combined influence. As such, their preferences (opinions) are expected to be very close to the average preferences (opinions).

In the following theorem, we present the main result of this paper by showing the percolation phenomenon in the competitive influence prorogation model. The proof of \rthe{main} is given in Appendix B.
\bthe{main}
Assume that
\begin{description}
\item[(i)] the initial PPV is uniformly distributed, i.e., $h_k=1/(b-1)$, $k=1, 2, \ldots, b-1$,
\item[(ii)] the mean field approximation for the combined influence on an undecided voter at time $t$ in \req{SBM9999} hold, and
\item[(iii)]
\beq{prefer2222}
p_{in}-\beta \frac{(\lambda_b)^2}{ \sum_{\ell=1}^b \rho_\ell \lambda_\ell} \ge 0.
\eeq
\end{description}
Let $k^*=\mathop{\arg\max}_{1 \le k \le b-1}z_k$, where
\bear{prefer1111}
&& z_{k}=\rho_k p_{out}+(\rho_{b,k} +{\rho^u_{b}}{h_k}) p_{in} \nonumber \\
&&\quad\quad-
\beta \frac{\rho_k \lambda_k \lambda_b+(\rho_{b,k}+{\rho^u_{b}}{h_k}) (\lambda_b)^2}{\sum_{\ell=1}^b \rho_\ell \lambda_\ell}.
\eear
Then for every undecided voter $u$, we have
\bear{prefer3333}
&&h_{u,k^*}(t) \ge h_{u,k}(t)\label{eq:prefer3333a},\\
&&z_{u,k^*}(t) \ge z_{u,k}(t)\label{eq:prefer3333b},\;\mbox{and}\\
&&h_{u,k^*}(t^+) \ge h_{u,k^*}(t) \label{eq:prefer3333c},
\eear
for all $t$ and $k \ne k^*$.
\ethe

\rthe{main} implies that if at time $0$ all the undecided voters have no particular preferences among the $b-1$ candidates and candidate ${k^*}$ is the candidate that has the largest combined influence on the undecided voters, then it remains the candidate that has the largest combined influence on the undecided voters at any time $t$. Moreover, it is also the most preferred candidate at any time $t$, and the preference of every undecided voter for candidate $k^*$ is increasing in time. As such, candidate $k^*$ is going to win over all the undecided votes at the ending time $T$. The physical meaning of the assumption in \req{prefer2222} is that the ``average'' influence between two undecided voters is nonnegative. Through nonnegative influence prorogation, candidate $k^*$ can receive higher preferences from undecided voters with respect to time. We also note that the assumption in (i) of \rthe{main} can be relaxed to the assumption that candidate $k^*$ is the most preferred candidate at time $0$, i.e., $h_{k^*} \ge h_k$ for all $k \ne k^*$.

In view of \rthe{main}, the strategy for candidate $k$ to win over all the undecided votes is to place $n \rho_{b,k}$ seeded voters in the set of undecided voters so that $z_k$ in \req{prefer1111} is larger than that of any other candidate. In other words, in order for a candidate to win over undecided votes, it needs to keep placing its seeded voters until the number of its seeded voters exceed a percolation threshold. Note that the percolation threshold depends on the numbers of seeded voters placed by the other candidates, the network parameters $p_{in}$ and $p_{out}$, and the social temperature $\beta$.

\bsec{Experimental results}{exp}

In this section, we perform various experiments to verify the performance and the percolation threshold of the competitive influence propagation model in Algorithm \ref{alg:propagation} by using the synthetic datasets generated from the stochastic block models and a real-world network.

\bsubsec{Stochastic block models}{expsbm}

\subsubsection{One eager candidate}\label{sec:one}

In this experiment, we consider the stochastic block model with $n=2000$, $b=3$, $p_{in} = 0.8$, $p_{out} = 0.2$, $\rho_1 = 0.5$, $\rho_2 = 0$, $\rho_3 = 0.5$, $\theta = 20$. As such, there are two candidates, i.e., $K = b-1 = 2$. Candidate 1 is eager to win the election and it already has $1000$ seeded voters in block 1. On the other hand, candidate 2 does not have any seeded voters. Undecided voters only exist in block 3. Suppose that candidate 1 would like to win over the votes from undecided voters and places additional $\rho_{3,1}\cdot n$ (with $\rho_{3,1} \le \rho_3=0.5$) seeded voters in block 3. On the other hand, candidate 2 places none of its seeded voters. As such,
\beq{exp0011}
\rho^u_b=\rho_3 -\rho_{3,1}=0.5-\rho_{3,1},
\eeq
and there are $n\rho^u_b$ undecided voters in block 3.

The question is then how many seeded voters candidate 1 need to place in block 3 in order to win over almost every undecided voter. To address such a question, we apply our percolation result in \rthe{main}. Note that in this setting
\bearn
&&\lambda_1=\lambda_3=\frac{p_{in}+ p_{out}}{2}=0.5, \\
&&\lambda_2=p_{out}=0.2.
\eearn
Thus,
$$\sum_{\ell=1}^3 \rho_l \lambda_l=0.5.$$
For the condition in \req{prefer2222} to hold, we need
\beq{exp1111}
p_{in}-\beta \frac{(\lambda_b)^2}{ \sum_{\ell=1}^b \rho_\ell \lambda_\ell} =0.8 -\beta \times 0.5 \ge 0.
\eeq
The initial PPV for an undecided voter is set to be $h=(h_1, h_2)=(0.5, 0.5)$ as that there is no particular preference for an undecided voter. Using \req{prefer1111}, we can compute
\bear{exp2222}
&&z_1=0.5 \times 0.2+(\rho_{3,1}+(0.5-\rho_{3,1}) \times 0.5) \times 0.8\nonumber\\
&&\quad-\beta \frac{0.5 \times 0.5 \times 0.5+(\rho_{3,1}+(0.5-\rho_{3,1}) \times 0.5) \times (0.5)^2}{0.5},\nonumber\\
&&z_2=(0.5-\rho_{3,1}) \times 0.5 \times 0.8\nonumber\\
&&\quad-\beta \frac{(0.5-\rho_{3,1}) \times 0.5) \times (0.5)^2}{0.5} .
\eear
In order for $z_1 \ge z_2$, candidate 1 needs to place $\rho_{3,1}$ seeded voters in block 3 with
\beq{exp3333}
\rho_{3,1} \ge \frac{1}{2} \frac{\beta-0.4}{1.6 -\beta}.
\eeq

\begin{figure}[tb]
	\centering
	\includegraphics[width=0.45\textwidth]{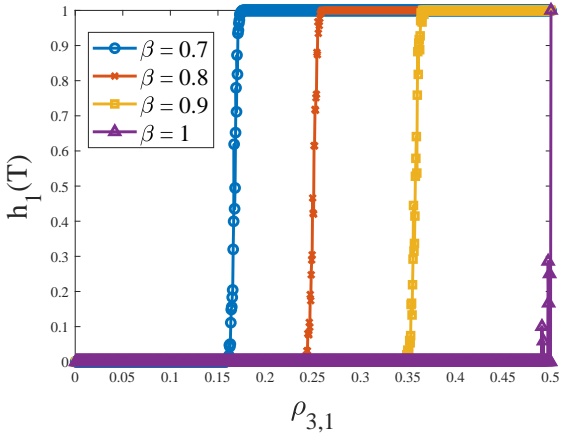}
	\caption{Percolation thresholds for one eager candidate}
	\label{fig:twocandidates1}
\end{figure}

In \rfig{twocandidates1}, we show the fraction of undecided voters who vote for candidate 1 at $T=10^6$ for $\beta = 0.7, 0.8, 0.9$, and $1$. The corresponding percolation thresholds for $\rho_{3,1}$ are $0.167$, $0.25$, $0.357$, and $0.5$, respectively. As shown in \rfig{twocandidates1}, these percolation thresholds for $\rho_{3,1}$ match extremely well with the simulation results.

\begin{figure}[tb]
	\centering
	\includegraphics[width=0.5\textwidth]{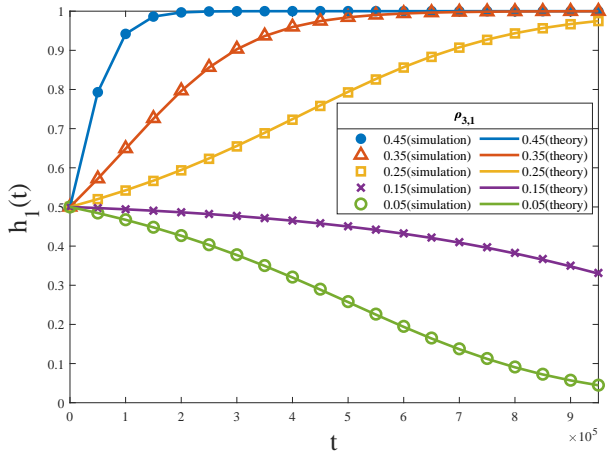}
	\caption{The average preference of undecided voters for candidate 1 over time, with $\beta = 0.7$ for $\rho_{3,1} = 0.05, 0.15, 0.25, 0.35, 0.45$, respectively. The theoretical results match extremely well with the simulation results.}
	\label{fig:twocandidates-time}
\end{figure}

In \rfig{twocandidates-time}, We show the average preference of undecided voters for candidate 1 over time, i.e., $h_1(t)$, with $\beta = 0.7$ for $\rho_{3,1} = 0.05, 0.15, 0.25, 0.35, 0.45$, respectively. Using \req{influence2222} and \req{SBM9999} yield theoretical results. Our theoretical results match extremely well with the simulation results. As shown in \rfig{twocandidates-time}, $h_1(t)$ increases to $1$ if $\rho_{3,1}$ exceeds the percolation threshold $0.167$. Moreover, the larger $\rho_{3,1}$ is, the faster the convergence is. On the other hand, it decreases to $0$ if $\rho_{3,1}$ is below the percolation threshold. 

\subsubsection{Two competing candidates}

The simulation setting is very similar to that in the previous section except that candidate 2 is also eager to win the election. To model this, we set $n = 3000$, $b = 4$, $\rho_1=\rho_2=\rho_4 =1/3$, $\rho_3 =0$, $p_{in} = 0.9$, $p_{out} = 0.1$, $\beta = 0.8$, $\theta = 100$. As such, there are three candidates, i.e., $K=b-1=3$. Candidate 1 and candidate 2 are competing to win the election, each of these two candidates already has $1000$ seeded voters in block 1 and block 2, respectively. On the other hand, we assume that the new candidate (candidate 3) does not have any seeded voter. Undecided voters only exist in block 4. Suppose that candidate 1 (resp. candidate 2) places additional $\rho_{4,1} \cdot n$ (resp. $\rho_{4,2} \cdot n$) (with $\rho_{4,1} \le \rho_4=1/3$) seeded voters in block 4 and that candidate 3 places none of its seeded voters in block 4. As such,
\beq{exp0022}
\rho^u_b=\rho_4 -\rho_{4,1} -\rho_{4,2}=1/3 -\rho_{4,1} -\rho_{4,2},
\eeq
and there are $n\rho^u_b$ undecided voters in block 4.

The question is then how many seeded voters candidate 1 (resp. candidate 2) need to place in block 4 in order to win over almost every undecided voter. To address such a question, we apply our percolation result in \rthe{main}. Note that in this setting
\bearn
&&\lambda_1=\lambda_2=\lambda_4=1/3 \times p_{in}+ 2/3 \times p_{out}=11/30, \\
&&\lambda_3=p_{out}=0.1.
\eearn
Thus,
$$\sum_{\ell=1}^4 \rho_l \lambda_l=11/30.$$
For the condition in \req{prefer2222} to hold, we need
\beq{exp2222a}
p_{in}-\beta \frac{(\lambda_b)^2}{ \sum_{\ell=1}^b \rho_\ell \lambda_\ell} =0.9 -\beta \times \frac{11}{30} \ge 0.
\eeq
The initial PPV for an undecided voter is set to be $h=(h_1, h_2, h_3)=(1/3, 1/3, 1/3)$ as that there is no particular preference for an undecided voter. Using \req{prefer1111}, we can compute
\bear{exp2222b}
&&z_1=\frac{1}{3} \times 0.1+(\rho_{4,1}+(\frac{1}{3}-\rho_{4,1}-\rho_{4,2}) \times \frac{1}{3}) \times 0.9\nonumber\\
&&\quad-\beta \frac{\frac{1}{3} \times \frac{11}{30}^2+(\rho_{4,1}+(\frac{1}{3}-\rho_{4,1}-\rho_{4,2}) \times \frac{1}{3}) \times \frac{11}{30}^2}{\frac{11}{30}},\nonumber\\
&&z_2=\frac{1}{3} \times 0.1+(\rho_{4,2}+(\frac{1}{3}-\rho_{4,1}-\rho_{4,2}) \times \frac{1}{3}) \times 0.9\nonumber\\
&&\quad-\beta \frac{\frac{1}{3} \times \frac{11}{30}^2+(\rho_{4,2}+(\frac{1}{3}-\rho_{4,1}-\rho_{4,2}) \times \frac{1}{3}) \times \frac{11}{30}^2}{\frac{11}{30}},\nonumber\\
&&z_3=(\frac{1}{3}-\rho_{4,1}-\rho_{4,2}) \times \frac{1}{3} \times 0.9\nonumber\\
&&\quad-\beta \frac{(\frac{1}{3}-\rho_{4,1}-\rho_{4,2}) \times \frac{1}{3} \times \frac{11}{30}^2}{\frac{11}{30}} .
\eear
In order for $z_1 \ge z_2$ and $z_1 \ge z_3$, candidate 1 needs to place $\rho_{4,1}$ seeded voters in block 3 with
\beq{exp0044}
\rho_{4,1} \ge \rho_{4,2} \ge \frac{\frac{33}{270} \times \beta-\frac{1}{30}}{0.9 -\frac{33}{90} \times \beta}.
\eeq
\begin{figure}[tb]
	\centering
	\includegraphics[width=0.45\textwidth]{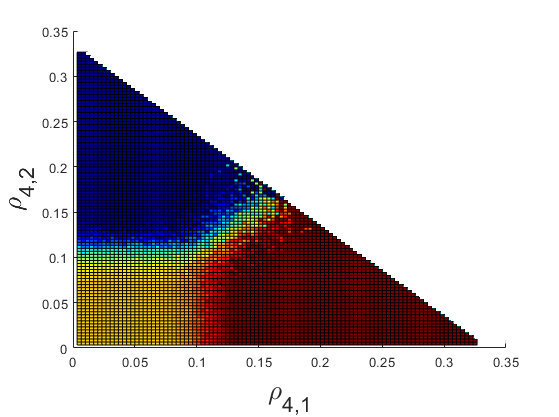}
	\caption{Percolation thresholds for two competing candidates}
	\label{fig:threecandidates1}
\end{figure}

In \rfig{threecandidates1}, we show the fraction of undecided voters who vote for these three candidates at $T=3 \times 10^6$ when $\beta = 0.8$. We use different colors to depict the final voting results: red for candidate 1, blue for candidate 2, and yellow for candidate 3. The darker the color of a candidate is, the larger fraction of votes for that candidate is. Clearly, as shown in \rfig{threecandidates1}, there are percolation thresholds for a candidate to win over the votes from undecided voters. For instance, if $\rho_{4,1} > \rho_{4,2}$ and $\rho_{4,1} > 0.106$, then candidate 1 wins (almost) all the votes from undecided voters. Once again, these percolation thresholds for $\rho_{4,1}$ match extremely well with the simulation results. On the other hand, if $\rho_{4,2} > \rho_{4,1}$ and $\rho_{4,2} > 0.106$, then candidate 2 wins (almost) all the votes from undecided voters. When $\rho_{4,1} < 0.106$ and $\rho_{4,2} < 0.106$, candidate 3 wins (almost) all the votes from undecided voters. The reason that candidate 3 can win all the votes from undecided voters in that setting is that we choose $\beta=0.8$ and the ``social temperature'' is high. Undecided voters who do not have links with seeded voters in blocks 1 and 2 tend to ``dislike'' these two candidates, and thus decide not to vote for them.

\bsubsec{Real-world network}{real}

\subsubsection{The network of Political Blogs}\label{sec:pol}

In this experiment, we evaluate our model on the real-world network ({\em Political Blogs}) in \cite{adamic2005political} that is obtained from the posts around the time of the United States presidential election of 2004. The original network consists of $1,490$ nodes and $19,090$ edges. After deleting nodes with degree less than or equal to $1$, we obtain a network with 1095 nodes and $16,587$ edges. For this network, we use their labels to partition the network into two blocks with $508$ nodes and $587$ nodes, respectively. The simulation setting is similar to that in \rsec{one}. For this dataset, we have $n=1095$, $b=3$, and the average intra-block edge probability $p_{in} = 0.9224$, the average inter-block edge probability $p_{out} = 0.0776$, $\rho_1=508/1095$, $\rho_2=0$, $\rho_3=587/1095$, $\theta = 20$. As such, there are two candidates, i.e., $K=b-1=2$. Candidate 1 is eager to win the election, and it already has $508$ seeded voters in block 1. On the other hand, candidate 2 does not have any seeded voters. Undecided voters only exist in block 3. Suppose that candidate 1 would like to win over the votes from undecided voters and places additional $\rho_{3,1} \cdot n$ (with $\rho_{3,1} \le \rho_3=587/1095$) seeded voters in block 3. On the other hand, candidate 2 places none of its seeded voters. As such,
\beq{exp0055}
\rho^u_b=\rho_3 -\rho_{3,1}=\frac{587}{1095}-\rho_{3,1},
\eeq
and there are $n\rho^u_b$ undecided voters in block 3.

The question is then how many seeded voters candidate 1 need to place in block 3 in order to win over almost every undecided voter. To address such a question, we apply our percolation result in \rthe{main}. Note that in this setting
\bearn
&&\lambda_1=\frac{508}{1095} \times 0.9224+\frac{587}{1095} \times 0.0776=0.469525, \\
&&\lambda_2=p_{out}=0.0776, \\
&&\lambda_3=\frac{587}{1095} \times 0.469525+\frac{508}{1095} \times 0.0776=0.530474.
\eearn
Thus,
$$\sum_{\ell=1}^3 \rho_l \lambda_l=0.502198.$$
For the condition in \req{prefer2222} to hold, we need
\beq{exp1111a}
p_{in}-\beta \frac{(\lambda_b)^2}{ \sum_{\ell=1}^b \rho_\ell \lambda_\ell} =0.9224 -\beta \times 0.56034 \ge 0.
\eeq
The initial PPV for an undecided voter is set to be $h=(h_1, h_2)=(0.5, 0.5)$ as that there is no particular preference for an undecided voter. Using \req{prefer1111}, we can compute
\bear{exp2222c}
&&z_1=\frac{508}{1095} \times 0.0776+(0.5 \times \rho_{3,1}+\frac{587}{2190}) \times 0.9224\nonumber\\
&&\quad-\beta \times ( \frac{508}{1095} \times 0.469525 \times 0.530474+\nonumber\\
&&\quad(\rho_{3,1}+(\frac{587}{1095}-\rho_{3,1}) \times 0.5) \times (0.530474)^2/0.5),\nonumber\\
&&z_2=(\frac{587}{1095}-\rho_{3,1}) \times 0.5 \times 0.9224\nonumber\\
&&\quad-\beta \frac{(\frac{587}{1095}-\rho_{3,1}) \times 0.5) \times (0.530474)^2}{0.502198} .
\eear
In order for $z_1 \ge z_2$, candidate 1 needs to place $\rho_{3,1}$ seeded voters in block 3 with
\beq{exp3333a}
\rho_{3,1} \ge \frac{0.23009 \times \beta-0.036}{\frac{1153}{1250} -0.56034 \times \beta}.
\eeq

\begin{figure}[tb]
	\centering
	\includegraphics[width=0.45\textwidth]{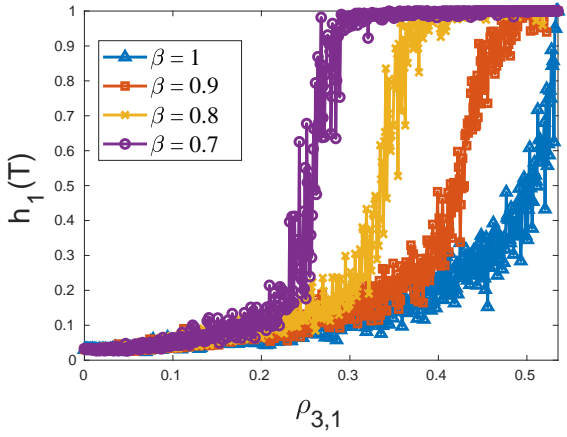}
	\caption{Percolation thresholds for Political Blogs}
	\label{fig:realdataset}
\end{figure}

In \rfig{realdataset}, we show the fraction of undecided voters who vote for candidate 1 at $T=10^6$ for $\beta = 0.7, 0.8, 0.9$, and $1$. The corresponding theoretical percolation thresholds for $\rho_{3,1}$ are $0.2358$, $0.3123$, $0.409$, and $0.536$, respectively. As shown in \rfig{realdataset}, we can also observe the phenomenon of percolation in this real-world network. However, there are roughly $5\thicksim10\%$ errors between the theoretical percolation thresholds and those estimated from the simulation results in \rfig{realdataset}.

\subsubsection{The Youtube dataset}\label{sec:youtube}

\begin{figure}[tb]
	\centering
	\includegraphics[width=0.45\textwidth]{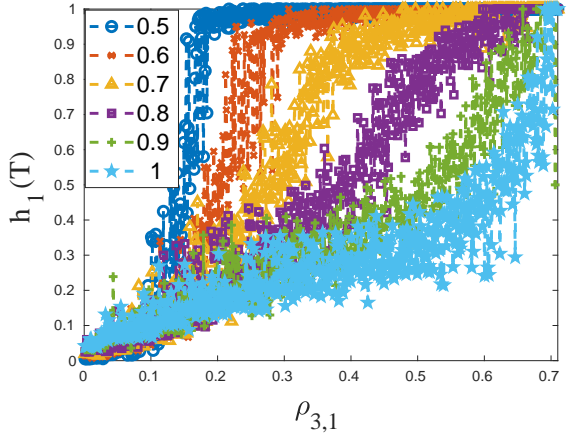}
	\caption{Percolation thresholds for the Youtube social network with two blocks}
	\label{fig:realdataset-youtube-2block}
\end{figure}

\begin{figure}[tb]
	\centering
	\includegraphics[width=0.45\textwidth]{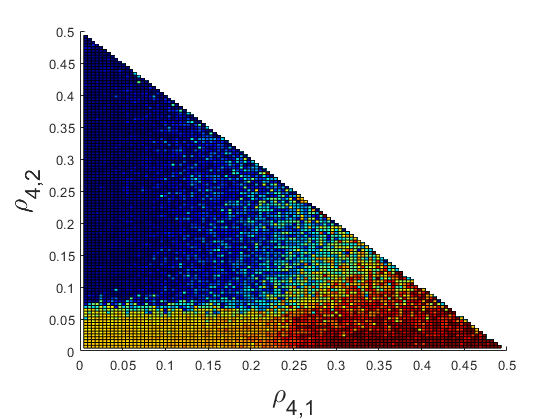}
	\caption{Percolation thresholds for the Youtube social network with three blocks}
	\label{fig:realdataset-youtube-3block}
\end{figure}

In this experiment, we evaluate our model on the Youtube social network in \cite{yang2015defining}. Youtube is a video-sharing web site that includes a social network. In the Youtube social network, if there is a connection between two users, they are friends. The original network consists of $1,134,890$ nodes and $2,987,624$ edges and $8,385$ overlapping communities in this dataset. We first convert overlapping communities into non-overlapping communities by using the maximum independent set method in \cite{nguyen2012containment}. As there are still too many communities, we only select the first two or three largest communities. Nodes with degrees smaller than two are deleted. Also, self-loop edges are deleted and multiple edges are replaced by a single edge so that the network is a {\em simple graph}. As a result, we obtain an undirected network containing two blocks of $798$ nodes (resp. three blocks with $900$ nodes). The experimental result for two (resp. three) blocks is shown in \rfig{realdataset-youtube-2block} (resp. \rfig{realdataset-youtube-3block}). We can also observe the percolation phenomenon in the Youtube dataset.

\subsubsection{The email dataset}\label{sec:email}

\begin{figure}[tb]
	\centering
	\includegraphics[width=0.45\textwidth]{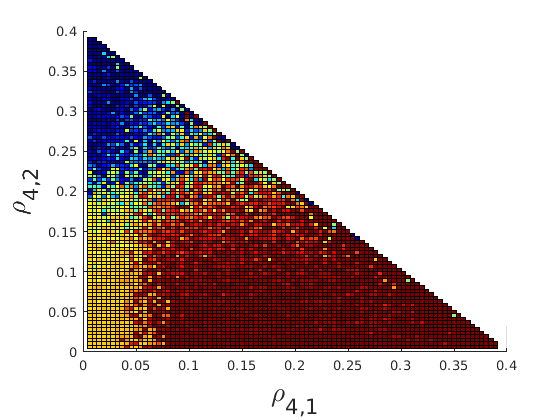}
	\caption{Percolation thresholds for the email dataset}
	\label{fig:realdataset-email}
\end{figure}

In this experiment, we evaluate our model for the email network in \cite{yin2017local, leskovec2007graph}. The email network was generated by using the email data from a large European research institution. If person $u$ sent person $v$ at least one email, then there is an edge between them. For this dataset, there are $1,005$ nodes and $25,571$ edges. For our experiments, we only select the first three largest groups. Nodes with degrees smaller than $1$ are deleted. Also, self-loop edges are deleted and multiple edges are replaced by a single edge so that the network is a {\em simple graph}. For this network, we use their labels to partition the network into three blocks with $52$ nodes, $86$ nodes, and $96$ nodes, respectively. As a result, we obtain the undirected network containing $230$ nodes with three blocks. The experimental result for this dataset is shown in \rfig{realdataset-email}. Clearly, we can still observe the percolation phenomenon.

\bsec{Conclusions}{con}

In this paper, we proposed a new influence propagation model for modeling the competitive influence of $K$ candidates among $n$ voters in an election process. In our influence propagation model, each voter is assigned with an initial $K$-dimensional probability preference vector (PPV). The PPV of a voter is then updated by using the softmax decision based on the combined influence. For such an influence propagation model, we showed that for a random network generated from the stochastic block model, there exists a percolation threshold for a candidate to win the election if the number of seeded voters placed by the candidate exceeds the threshold. By conducting extensive experiments, we also showed that our theoretical percolation thresholds are very close to those obtained from simulations for random networks and the errors are within $10\%$ for the network of Political Blogs. For the Youtube dataset and the email dataset, our simulation results also show the percolation phenomenon.
Our future work is to further refine our model by considering more general random networks, such as the configuration model \cite{Newman2010}. In such a general network model, a candidate can also choose a set of seeded voters that might depend on the degrees of nodes to maximize its influence.

\bibliographystyle{IEEEtran}
\bibliography{refs}

\begin{thebibliography}{10}
\providecommand{\url}[1]{#1}
\csname url@samestyle\endcsname
\providecommand{\newblock}{\relax}
\providecommand{\bibinfo}[2]{#2}
\providecommand{\BIBentrySTDinterwordspacing}{\spaceskip=0pt\relax}
\providecommand{\BIBentryALTinterwordstretchfactor}{4}
\providecommand{\BIBentryALTinterwordspacing}{\spaceskip=\fontdimen2\font plus
\BIBentryALTinterwordstretchfactor\fontdimen3\font minus
  \fontdimen4\font\relax}
\providecommand{\BIBforeignlanguage}[2]{{%
\expandafter\ifx\csname l@#1\endcsname\relax
\typeout{** WARNING: IEEEtran.bst: No hyphenation pattern has been}%
\typeout{** loaded for the language `#1'. Using the pattern for}%
\typeout{** the default language instead.}%
\else
\language=\csname l@#1\endcsname
\fi
#2}}
\providecommand{\BIBdecl}{\relax}
\BIBdecl

\bibitem{kempe2003maximizing}
D.~Kempe, J.~Kleinberg, and {\'E}.~Tardos, ``Maximizing the spread of influence
  through a social network,'' in \emph{Proceedings of the ninth ACM SIGKDD
  international conference on Knowledge discovery and data mining}.\hskip 1em
  plus 0.5em minus 0.4em\relax ACM, 2003, pp. 137--146.

\bibitem{bharathi2007competitive}
S.~Bharathi, D.~Kempe, and M.~Salek, ``Competitive influence maximization in
  social networks,'' in \emph{International workshop on web and internet
  economics}.\hskip 1em plus 0.5em minus 0.4em\relax Springer, 2007, pp.
  306--311.

\bibitem{carnes2007maximizing}
T.~Carnes, C.~Nagarajan, S.~M. Wild, and A.~Van~Zuylen, ``Maximizing influence
  in a competitive social network: a follower's perspective,'' in
  \emph{Proceedings of the ninth international conference on Electronic
  commerce}.\hskip 1em plus 0.5em minus 0.4em\relax ACM, 2007, pp. 351--360.

\bibitem{kostka2008word}
J.~Kostka, Y.~A. Oswald, and R.~Wattenhofer, ``Word of mouth: Rumor
  dissemination in social networks,'' in \emph{International colloquium on
  structural information and communication complexity}.\hskip 1em plus 0.5em
  minus 0.4em\relax Springer, 2008, pp. 185--196.

\bibitem{he2012influence}
X.~He, G.~Song, W.~Chen, and Q.~Jiang, ``Influence blocking maximization in
  social networks under the competitive linear threshold model,'' in
  \emph{Proceedings of the 2012 siam international conference on data
  mining}.\hskip 1em plus 0.5em minus 0.4em\relax SIAM, 2012, pp. 463--474.

\bibitem{shirazipourazad2012influence}
S.~Shirazipourazad, B.~Bogard, H.~Vachhani, A.~Sen, and P.~Horn, ``Influence
  propagation in adversarial setting: how to defeat competition with least
  amount of investment,'' in \emph{Proceedings of the 21st ACM international
  conference on Information and knowledge management}.\hskip 1em plus 0.5em
  minus 0.4em\relax ACM, 2012, pp. 585--594.

\bibitem{lin2015learning}
S.-C. Lin, S.-D. Lin, and M.-S. Chen, ``A learning-based framework to handle
  multi-round multi-party influence maximization on social networks,'' in
  \emph{Proceedings of the 21th ACM SIGKDD International Conference on
  Knowledge Discovery and Data Mining}.\hskip 1em plus 0.5em minus 0.4em\relax
  ACM, 2015, pp. 695--704.

\bibitem{lin2015analyzing}
Y.~Lin and J.~C. Lui, ``Analyzing competitive influence maximization problems
  with partial information: An approximation algorithmic framework,''
  \emph{Performance Evaluation}, vol.~91, pp. 187--204, 2015.

\bibitem{li2015getreal}
H.~Li, S.~S. Bhowmick, J.~Cui, Y.~Gao, and J.~Ma, ``Getreal: Towards realistic
  selection of influence maximization strategies in competitive networks,'' in
  \emph{Proceedings of the 2015 ACM SIGMOD international conference on
  management of data}.\hskip 1em plus 0.5em minus 0.4em\relax ACM, 2015, pp.
  1525--1537.

\bibitem{tong2018misinformation}
A.~Tong, D.-Z. Du, and W.~Wu, ``On misinformation containment in online social
  networks,'' in \emph{Advances in Neural Information Processing Systems},
  2018, pp. 339--349.

\bibitem{tong2018distributed}
G.~Tong, W.~Wu, and D.-Z. Du, ``Distributed rumor blocking with multiple
  positive cascades,'' \emph{IEEE Transactions on Computational Social
  Systems}, vol.~5, no.~2, pp. 468--480, 2018.

\bibitem{das2014modeling}
A.~Das, S.~Gollapudi, and K.~Munagala, ``Modeling opinion dynamics in social
  networks,'' in \emph{Proceedings of the 7th ACM international conference on
  Web search and data mining}.\hskip 1em plus 0.5em minus 0.4em\relax ACM,
  2014, pp. 403--412.

\bibitem{degroot1974reaching}
M.~H. DeGroot, ``Reaching a consensus,'' \emph{Journal of the American
  Statistical Association}, vol.~69, no. 345, pp. 118--121, 1974.

\bibitem{hegselmann2002opinion}
R.~Hegselmann, U.~Krause \emph{et~al.}, ``Opinion dynamics and bounded
  confidence models, analysis, and simulation,'' \emph{Journal of artificial
  societies and social simulation}, vol.~5, no.~3, 2002.

\bibitem{clifford1973model}
P.~Clifford and A.~Sudbury, ``A model for spatial conflict,''
  \emph{Biometrika}, vol.~60, no.~3, pp. 581--588, 1973.

\bibitem{holley1975ergodic}
R.~A. Holley, T.~M. Liggett \emph{et~al.}, ``Ergodic theorems for weakly
  interacting infinite systems and the voter model,'' \emph{The annals of
  probability}, vol.~3, no.~4, pp. 643--663, 1975.

\bibitem{proskurnikov2016opinion}
A.~V. Proskurnikov, A.~S. Matveev, and M.~Cao, ``Opinion dynamics in social
  networks with hostile camps: Consensus vs. polarization,'' \emph{IEEE
  Transactions on Automatic Control}, vol.~61, no.~6, pp. 1524--1536, 2016.

\bibitem{dhamal2018two}
S.~Dhamal, W.~Ben-Ameur, T.~Chahed, and E.~Altman, ``A two phase investment
  game for competitive opinion dynamics in social networks,'' \emph{arXiv
  preprint arXiv:1811.08291}, 2018.

\bibitem{morone2015influence}
F.~Morone and H.~A. Makse, ``Influence maximization in complex networks through
  optimal percolation,'' \emph{Nature}, vol. 524, no. 7563, p.~65, 2015.

\bibitem{adamic2005political}
L.~A. Adamic and N.~Glance, ``The political blogosphere and the 2004 us
  election: divided they blog,'' in \emph{Proceedings of the 3rd international
  workshop on Link discovery}.\hskip 1em plus 0.5em minus 0.4em\relax ACM,
  2005, pp. 36--43.

\bibitem{yang2015defining}
J.~Yang and J.~Leskovec, ``Defining and evaluating network communities based on
  ground-truth,'' \emph{Knowledge and Information Systems}, vol.~42, no.~1, pp.
  181--213, 2015.

\bibitem{yin2017local}
H.~Yin, A.~R. Benson, J.~Leskovec, and D.~F. Gleich, ``Local higher-order graph
  clustering,'' in \emph{Proceedings of the 23rd ACM SIGKDD International
  Conference on Knowledge Discovery and Data Mining}.\hskip 1em plus 0.5em
  minus 0.4em\relax ACM, 2017, pp. 555--564.

\bibitem{leskovec2007graph}
J.~Leskovec, J.~Kleinberg, and C.~Faloutsos, ``Graph evolution: Densification
  and shrinking diameters,'' \emph{ACM Transactions on Knowledge Discovery from
  Data (TKDD)}, vol.~1, no.~1, p.~2, 2007.

\bibitem{xiao2007distributed}
L.~Xiao, S.~Boyd, and S.-J. Kim, ``Distributed average consensus with
  least-mean-square deviation,'' \emph{Journal of parallel and distributed
  computing}, vol.~67, no.~1, pp. 33--46, 2007.

\bibitem{boyd2006randomized}
S.~Boyd, A.~Ghosh, B.~Prabhakar, and D.~Shah, ``Randomized gossip algorithms,''
  \emph{IEEE/ACM Transactions on Networking (TON)}, vol.~14, no.~SI, pp.
  2508--2530, 2006.

\bibitem{gold1996softmax}
S.~Gold, A.~Rangarajan \emph{et~al.}, ``Softmax to softassign: Neural network
  algorithms for combinatorial optimization,'' \emph{Journal of Artificial
  Neural Networks}, vol.~2, no.~4, pp. 381--399, 1996.

\bibitem{bishop2006pattern}
C.~M. Bishop, \emph{Pattern recognition and machine learning}.\hskip 1em plus
  0.5em minus 0.4em\relax springer, 2006.

\bibitem{goyal2010learning}
A.~Goyal, F.~Bonchi, and L.~V. Lakshmanan, ``Learning influence probabilities
  in social networks,'' in \emph{Proceedings of the third ACM international
  conference on Web search and data mining}.\hskip 1em plus 0.5em minus
  0.4em\relax ACM, 2010, pp. 241--250.

\bibitem{liao2016uncovering}
P.-L. Liao, C.-K. Chou, and M.-S. Chen, ``Uncovering multiple diffusion
  networks using the first-hand sharing pattern,'' in \emph{Proceedings of the
  2016 SIAM International Conference on Data Mining}.\hskip 1em plus 0.5em
  minus 0.4em\relax SIAM, 2016, pp. 63--71.

\bibitem{reichardt2006statistical}
J.~Reichardt and S.~Bornholdt, ``Statistical mechanics of community
  detection,'' \emph{Physical Review E}, vol.~74, no.~1, p. 016110, 2006.

\bibitem{Newman04}
M.~E. Newman, ``Fast algorithm for detecting community structure in networks,''
  \emph{Physical review E}, vol.~69, no.~6, p. 066133, 2004.

\bibitem{chang2015relative}
C.-S. Chang, C.-J. Chang, W.-T. Hsieh, D.-S. Lee, L.-H. Liou, and W.~Liao,
  ``Relative centrality and local community detection,'' \emph{Network
  Science}, vol.~3, no.~4, pp. 445--479, 2015.

\bibitem{chang2018probabilistic}
C.-S. Chang, D.-S. Lee, L.-H. Liou, S.-M. Lu, and M.-H. Wu, ``A probabilistic
  framework for structural analysis and community detection in directed
  networks,'' \emph{IEEE/ACM Transactions on Networking (TON)}, vol.~26, no.~1,
  pp. 31--46, 2018.

\bibitem{erdos1959random}
P.~Erdos, ``On random graphs,'' \emph{Publicationes mathematicae}, vol.~6, pp.
  290--297, 1959.

\bibitem{saade2014spectral}
A.~Saade, F.~Krzakala, and L.~Zdeborov{\'a}, ``Spectral clustering of graphs
  with the bethe hessian,'' in \emph{Advances in Neural Information Processing
  Systems}, 2014, pp. 406--414.

\bibitem{decelle2012mode}
L.~A. Decelle, F.~Krzakala, and P.~Zhang, ``Mode-net: Modules detection in
  networks,'' 2012.

\bibitem{nguyen2012containment}
N.~P. Nguyen, G.~Yan, M.~T. Thai, and S.~Eidenbenz, ``Containment of
  misinformation spread in online social networks,'' in \emph{Proceedings of
  the 4th Annual ACM Web Science Conference}.\hskip 1em plus 0.5em minus
  0.4em\relax ACM, 2012, pp. 213--222.

\bibitem{Newman2010}
M.~Newman, \emph{Networks: an introduction}.\hskip 1em plus 0.5em minus
  0.4em\relax OUP Oxford, 2009.

\end{thebibliography}


\begin{thebibliography}{99}
\bibitem{b1} P. Domingos and M. Richardson, ``Mining the network value of customers,'' in Proceedings of the 7th ACM SIGKDD Conference on Knowledge Discovery and Data Mining, 2001, pp. 57-66.
\bibitem{b2} M. Richardson and P. Domingos, ``Mining knowledge-sharing sites for viral marketing,'' in Proceedings of the 8th ACM SIGKDD Conference on Knowledge Discovery and Data Mining, 2002, pp.61-70.
\bibitem{b3} D. Kempe, J. M. Kleinberg, and E. Tardos, ``Maximizing the Spread of Influence through a Social Network,'' in KDD, 2003.
\bibitem{b4} Wei Chen, Yajun Wang, and Siyu Yang, ``Efficient Influence maximization in Social Networks,'' in Proceedings of the 15th ACM SIGKDD, 2009.
\bibitem{b5} A. Coja-Oghlan, U. Feige, M. Krivelevich, and D. Reichman ``Contagious Sets in Expanders,'' in Proceedings of the twenty-sixth annual ACM-SIAM symposium on discrete algorithms, 2015.
\bibitem{b6} S. Khanna, and B. Lucier ``Influence maximization in undirected networks,'' in Proceedings of the twenty-fifth annual ACM-SIAM symposium on Discrete algorithms, 2014.
\bibitem{b7} W. Chen, T. Lin, Z. Tan, M. Zhao, and X. Zhou, ``Robust influence maximization,'' in Proceedings of the 22nd ACM SIGKDD International Conference on Knowledge Discovery and Data Mining, 2016.
\bibitem{b8} Y. Tang, X. Xiao, and Y. Shi, ``Influence Maximization: Near-Optimal Time Complexity Meets Practical Efficiency,'' in Network Science, vol. 3, no. 4 , pp.445-479, 2015.
\bibitem{b9} S. Bharathi, D. Kempe, and M. Salek, ``Competitive influence maximization in social networks,'' in International Workshop on Web and Internet Economics, 2007.
\bibitem{b10} J. Kostka, Y. A. Oswald, and R.Wattenhofer, ``Word of mouth:
Rumor dissemination in social networks,'' in SIROCCO, pages 185–196, 2008.
\bibitem{b11} X. He, G. Song, W. Chen, Q. Jiang, ``Influence Blocking Maximization in Social Networks under the Competitive Linear Threshold Model,'' in Society for Industrial and Applied Mathematics, 2012.
\bibitem{b12} Lin, Su-Chen, Shou-De Lin, and Ming-Syan Chen, ``A Learning-based Framework to Handle Multi-round Multi-party Influence Maximization on Social Networks,'' in Proceedings of the 21th ACM SIGKDD International Conference on Knowledge Discovery and Data Mining. ACM, 2015.
\bibitem{b13}Tong, A., Du, D. Z., and Wu, W., ``On Misinformation Containment in Online Social Networks,'' In Advances in Neural Information Processing Systems (pp. 339-349), 2018.
\bibitem{b14} Tong, Guangmo, Weili Wu, and Ding-Zhu Du, ``Distributed Rumor Blocking With Multiple Positive Cascades,'' IEEE Transactions on Computational Social Systems 5.2 (2018): 468-480.
\bibitem{b15} T. Carnes, C. Nagarajan, S. M. Wild, A. V. Zuylen, ``Maximizing influence in a competitive social network: a follower's perspective,'' in Proceedings of the ninth international conference on Electronic commerce, 2007.
\bibitem{b16} H. Li, S. S. Bhowmick, J. Cui, Y. Gao, and J. Ma, ``Getreal: Towards realistic selection of influence maximization strategies in competitive networks,'' in Proceedings of the 2015 ACM SIGMOD international conference on management of data, 2015.
\bibitem{b17} S. Shirazipourazad, B. Bogard, H. Vachhani, A. Sen, and P. Horn, ``Influence propagation in adversarial setting: how to defeat competition with least amount of investment,'' in Proceedings of the 21st ACM international conference on Information and knowledge management, 2012.
\bibitem{b18} Lin Xiao, Stephen Boyd, Seung-Jean Kim, ``Distributed average consensus with least-mean-square deviation,'' in Journal of parallel and distributed computing, 2007.
\bibitem{b19} Stephen Boyd, Arpita Ghosh, Balaji Prabhakar, and Devavrat Shah, ``Randomized gossip algorithms,'' in IEEE Transactions on Information Theory, VOL. 52, NO. 6, pp. 2508-2530, JUNE 2006.
\bibitem{b20a} S. Gold, A. Rangarajan ``Softmax to softassign: neural network algorithms for combinatorial optimization,''
{\em Journal of Artificial Neural Networks}, vol. 2, no. 4, pp. 381--399. 1996.
\bibitem{b20} C. M. Bishop, ``Pattern recognition,'' Machine Learning, vol. 128, pp. 1-58, 2006.
\bibitem{b21} Jörg Reichardt, and Stefan Bornholdt, ``Statistical mechanics of community detection,'' in Physical Review E, 2006.
\bibitem{b22} C.-S. Chang, C.-J. Chang, W.-T. Hsieh, D.-S. Lee, L.-H. Liou, and W. Liao, ``Relative centrality and local community detection,'' in IEEE/ACM Transactions on Networking, 2017.
\bibitem{b23} C.-S. Chang, D.-S. Lee, L.-H. Liou, S.-M. Lu, and M.-H.Wu, ``A probabilistic framework for structural analysis and community detection in directed networks,'' in IEEE/ACM
Transactions on Networking, 2017.
\bibitem{Newman04}
M.~E. Newman, ``Fast algorithm for detecting community structure in networks,''
 \emph{Phys. Rev. E}, vol.~69, no.~6, p. 066133, 2004.

\bibitem{b24} P. Erdős and A. Rényi, ``On random graphs,'' Publ. Math. Debrecen,
vol. 6, pp. 290–297, 1959.
\bibitem{b25} A. Saade, F. Krzakala, and L. Zdeborova, ``Spectral clustering of graphs with the bethe hessian,'' in Proc. Ann. Conf. Advances in Neural Information Processing Systems, 2014, pp. 406–414.
\bibitem{b26} L. Z. Aurelien Decelle, Florent Krzakala and P. Zhang. (2012)
Mode-net: Modules detection in networks. [Online]. Available: http : //modenet.krzakala.org/


\end{thebibliography}

\newpage
\appendices
\section{Proof of \rlem{SBM0}}\label{lemma1}
\setcounter{section}{1}

Consider an $\mbox{SBM}(n,b,p_{in}, p_{out},\rho)$ with the $n \times n$ adjacency matrix $A=(a_{u,w})$ and a node $v$ is in block $i$. From the construction of the stochastic block model, we know that for $j \ne i$, $\sum_{w \in B_j} a_{u,w}$ is a sum of independent Bernoulli random variables with the parameter $p_{out}$. We then have from the strong law of large numbers that
\beq{SBM3366}
\lim_{n \to \infty}{\frac{\sum_{w \in B_j} a_{u,w}}{n}}=\rho_j p_{out}, \quad a.s.
\eeq
Similarity,
\beq{SBM3377}
\lim_{n \to \infty}{\frac{\sum_{w \in B_i} a_{u,w}}{n}}=\rho_i p_{in}, \quad a.s.
\eeq
Let $k_v=\sum_{w=1}^n a_{u,w}$ be the degree of node $v$. In view of \req{SBM3366} and \req{SBM3377}, we have
\bear{SBM3300a}
\lim_{n \to \infty}{\frac{k_v}{n}}&=&\rho_i p_{in}+\sum_{j \ne i}\rho_j p_{out}\nonumber\\
&=&\rho_i p_{in}+(1-\rho_i)p_{out}, \quad a.s. \nonumber \\
&=&\lambda_i,\quad a.s.
\eear
where $\lambda_i$ is defined in \req{SBM3355}.
For the ease of our representation, we represent the limit in \req{SBM3300a} by using the following simplified notation:
\beq{SBM3300}
{\frac{k_v}{n}}=\lambda_i+o(1).
\eeq
Note that $\lambda_i$ can be viewed as the normalized degree of a node in block $i$

Let $m$ be the total number of edges in the random graph and $B_i$, $i=1, 2, \ldots, b$, be the set of nodes in block $i$. Since every edge has two ends, it follows that
\beq{SBM3311}
2m =\sum_{v=1}^n k_v=\sum_{i=1}^b \sum_{v \in B_i} k_v.
\eeq
In conjunction with \req{SBM3300}, we then have
\beq{SBM3322}
\frac{2m}{n^2}=\sum_{i=1}^b \rho_i \lambda_i +o(1).
\eeq

Recall from \req{influence1111} that for an undecided voter $u$ in $B^u_b$,
\bear{SBM7777}
&&z_{u,k}(t)=\sum_{w \ne u}q(w,u)h_{w,k}(t) \nonumber\\
&&=\sum_{i=1}^{b-1}\sum_{w \in B_i}q(w,u)h_{w,k}(t)+\sum_{w \ne u, w \in B_b}q(w,u)h_{w,k}(t) \nonumber\\
&&=\sum_{w \in B_k}q(w,u)+\sum_{w \ne u, w \in B_b}q(w,u)h_{w,k}(t),
\eear
where we use $h_{w,k}(t)=\delta_{i,k}$ for $w \in B_i$ in the last identity. For the generalized modularity in
\req{modularity5555}, we have
\beq{SBM8888}
\sum_{w \in B_k}q(w,u)=\sum_{w \in B_k}\Big (\frac{a_{u,w}}{2m} - \beta \frac{k_w}{2m}\frac{k_u}{2m} \Big) .
\eeq
Using \req{SBM3366}, \req{SBM3300} and \req{SBM3322} in \req{SBM8888} yields
\bear{SBM8800}
&&n\sum_{w \in B_k}q(w,u)=\Big (\frac{\sum_{w \in B_k}a_{u,w}}{n} \Big) (\frac{n^2}{2m})\nonumber \\
 &&\quad\quad\quad - \beta \frac{1}{n}\sum_{w \in B_k}(\frac{k_w}{n})(\frac{n^2}{2m})(\frac{k_u}{n})(\frac{n^2}{2m}) \nonumber \\
&=&\frac{\rho_k}{\sum_{\ell=1}^b \rho_\ell \lambda_\ell}\Big (p_{out}-\beta \frac{\lambda_k \lambda_b}{ \sum_{\ell=1}^b \rho_\ell \lambda_\ell} \Big )+o(1).
\eear
Similarly, we can further decompose the second sum in \req{SBM7777} as follows:
\bear{SBM8811}
&&\sum_{w \ne u, w \in B_b}q(w,u)h_{w,k}(t) \nonumber \\
&&=\sum_{i=1}^{b-1}\sum_{w \in B_{b,i}}q(w,u)h_{w,k}(t)+\sum_{w \ne u, w \in B^u_b}q(w,u)h_{w,k}(t) \nonumber \\
&&=\sum_{w \in B_{b,k}}q(w,u)+\sum_{w \ne u, w \in B^u_b}q(w,u)h_{w,k}(t).
\eear
Using \req{SBM3377}, \req{SBM3300} and \req{SBM3322}, we have
\bear{SBM8822}
&&n \sum_{w \in B_{b,k}}q(w,u)=\Big (\frac{\sum_{w \in B_{b,k}}a_{u,w}}{n} \Big) (\frac{n^2}{2m})\nonumber \\
 &&\quad\quad\quad - \beta \frac{1}{n}\sum_{w \in B_{b,k}}(\frac{k_w}{n})(\frac{n^2}{2m})(\frac{k_u}{n})(\frac{n^2}{2m}) \nonumber \\
&&
= \frac{\rho_{b,k}}{\sum_{\ell=1}^b \rho_\ell \lambda_\ell}\Big (p_{in}-\beta \frac{(\lambda_b)^2}{ \sum_{\ell=1}^b \rho_\ell \lambda_\ell} \Big )+o(1).
\eear
Similarly,
\bear{SBM8833a}
&& n\sum_{w \ne u, w \in B^u_b}q(w,u)h_{w,k}(t)\nonumber \\
&&=\Big (\frac{\sum_{w \ne u,w \in B^u_b}a_{u,w}h_{w,k}(t)}{n} \Big) (\frac{n^2}{2m})\nonumber \\
 &&\; - {\beta} \frac{1}{n}\sum_{w \ne u, w \in B^u_b}h_{w,k}(t) (\frac{k_w}{n})(\frac{n^2}{2m})(\frac{k_u}{n})(\frac{n^2}{2m})+o(1). \nonumber \\
\eear
Since $\{a_{u,w}, w \in B^u_b\}$ are independent Bernoulli random variables with mean $p_{in}$, we use the mean field approximation to approximate the weighted sum of independent random variables ${\sum_{w \ne u, w \in B^u_b}a_{u,w}h_{w,k}(t)}/{n}$ by its mean as follows:
\bear{SBM8831}
&&\frac{\sum_{w \ne u, w \in B^u_b}a_{u,w}h_{w,k}(t)}{n} \nonumber \\
&&\approx\frac{\sum_{w \ne u,w \in B^u_b}p_{in}h_{w,k}(t)}{n}\nonumber\\
&&=\rho^u_{b} p_{in} h_k(t)+o(1).
\eear
Using \req{SBM8831}, \req{SBM3300} and \req{SBM3322} in \req{SBM8833a} yields
\bear{SBM8833}
&& n\sum_{w \ne u, w \in B^u_b}q(w,u)h_{w,k}(t)
 \nonumber \\
&&
 \approx\frac{\rho^u_{b}h_k(t)}{\sum_{\ell=1}^b \rho_\ell \lambda_\ell}\Big (p_{in}-\beta \frac{(\lambda_b)^2}{ \sum_{\ell=1}^b \rho_\ell \lambda_\ell} \Big ).
\eear
Using \req{SBM8800}, \req{SBM8822}, and \req{SBM8833} in \req{SBM7777} yields \req{SBM9999}.

\section{Proof of \rthe{main}}\label{theorem}

We prove \rthe{main} by induction on $t$. For $t=0$, the inequality in \req{prefer3333a} holds trivially at $t=0$ as $h_{u,k}(0)=1/(b-1)$ for all $k$. Also, we have from \req{SBM9999} in \rlem{SBM0} and \req{prefer1111} that $z_{u,k^*}(0) \ge z_{u,k}(0)$ for all $k \ne k^*$ and the inequality in \req{prefer3333b} also holds at time $0$.

Now suppose that there is a clock tick of the undecided voter $u$ at time $t$. Then the PPV of voter $u$ is updated according to \req{influence2222}. This implies that
\bear{prefer4444}
\frac{h_{u,k^*}(t^+)}{h_{u,k}(t^+)}&=&\frac{e^{\theta z_{u,k^*}(t)} h_{u,k^*}(t)}{e^{\theta z_{u,k}(t)} h_{u,k}(t)} \nonumber \\
&=&e^{\theta (z_{u,k^*}(t)-z_{u,k}(t))} \frac{h_{u,k^*}(t)}{h_{u,k}(t)}.
\eear
Since $z_{u,k^*}(t)\ge z_{u,k}(t)$ (from the induction hypothesis in \req{prefer3333b}) and $\theta \ge 0$, we have
\beq{prefer5555}
\frac{h_{u,k^*}(t^+)}{h_{u,k}(t^+)}\ge
\frac{h_{u,k^*}(t)}{h_{u,k}(t)}.
\eeq
In conjunction with the induction hypothesis in \req{prefer3333a}, i.e., ${h_{u,k^*}(t)} \ge {h_{u,k}(t)}$, it then follows that
${h_{u,k^*}(t^+)} \ge {h_{u,k}(t^+)}$.

To show that the induction hypothesis in \req{prefer3333c} holds, we rewrite\req{prefer5555} as follows:
\beq{prefer5555b}
{h_{u,k^*}(t^+)}{h_{u,k}(t)}\ge
{h_{u,k^*}(t)}{h_{u,k}(t^+)}.
\eeq
Since PPVs are probability distributions, summing over $k$ on both sides of \req{prefer5555b} yields
\beq{prefer5566}
{h_{u,k^*}(t^+)} \ge {h_{u,k^*}(t)}.
\eeq

It remains to show that the induction hypothesis $z_{v,k^*}(t^+) \ge z_{v,k}(t^+)$ in \req{prefer3333b} for every undecided voter $v$ after the update of the PPV of voter $u$ at time $t$. Using the mean field approximation in \req{SBM9999} yields
\bear{prefer6666}
&&n\Big (z_{v,k^*}(t^+) - z_{v,k}(t^+)-(z_{v,k^*}(t) - z_{v,k}(t)) \Big) \nonumber\\
&&\approx \frac{\rho^u_{b}}{\sum_{\ell=1}^b \rho_\ell \lambda_\ell}\Big (p_{in}-\beta \frac{(\lambda_b)^2}{ \sum_{\ell=1}^b \rho_\ell \lambda_\ell}\Big) \nonumber \\
&&\quad\Big (h_{k^*}(t^+)-h_k(t^+)-h_{k^*}(t)+h_k(t) \Big).
\eear
Since only the PPV of voter $u$ is updated at time $t$, we have from \req{SBM9955} and the update rule in \req{influence2222} that
\bear{prefer7777}
&&h_{k^*}(t^+)-h_k(t^+)-h_{k^*}(t)+h_k(t) \nonumber\\
&&=\frac{1}{n\rho^u_{b}}\Big (h_{u,k^*}(t^+)-h_{u,k}(t^+)-h_{u,k^*}(t)+h_{u,k}(t) \Big ) \nonumber \\
&&=\frac{1}{n\rho^u_{b}}\Big ((ce^{\theta z_{u,k^*}(t)}-1)h_{u,k^*}(t)\nonumber\\
&&\quad\quad-(ce^{\theta z_{u,k}(t)}-1)h_{u,k}(t) \Big ),
\eear
where $c=1/\sum_{\ell=1}^{b-1}e^{\theta z_{u,\ell}(t)}h_{u, \ell}(t)$ is the normalization constant. Now we show that
\beq{prefer8888}
(ce^{\theta z_{u,k^*}(t)}-1)h_{u,k^*}(t)-(ce^{\theta z_{u,k}(t)}-1)h_{u,k}(t) \ge 0.
\eeq
Since we have shown in \req{prefer5566} that ${h_{u,k^*}(t^+)} \ge {h_{u,k^*}(t)}$, we know that $ce^{\theta z_{u,k^*}(t)}-1 \ge 0$. If $ce^{\theta z_{u,k}(t)}-1<0$, then the inequality in \req{prefer8888} holds trivially. On the other hand, if $ce^{\theta z_{u,k}(t)}-1 \ge 0$, then we have from the induction hypotheses $z_{u,k^*}(t) \ge z_{u,k}(t)$ and $h_{u,k^*}(t) \ge h_{u,k}(t)$ that the inequality in \req{prefer8888} also holds. Thus,
\beq{prefer8855}
h_{k^*}(t^+)-h_k(t^+)-h_{k^*}(t)+h_k(t)\ge 0.
\eeq
Using \req{prefer8855} and \req{prefer2222} in \req{prefer6666} yields
\beq{prefer9999}
z_{v,k^*}(t^+) - z_{v,k}(t^+) - (z_{v,k^*}(t) - z_{v,k}(t)) \ge 0.
 \eeq
From the induction hypothesis $z_{v,k^*}(t) \ge z_{v,k}(t)$, we then have $z_{v,k^*}(t^+) \ge z_{v,k}(t^+)$. This then concludes the proofs for all the three induction hypotheses.

\end{document}